\documentclass[12pt]{article}
\begin{document}

\title{\bf  Symmetrized Schr\"{o}dinger Equation and General Complex Solution
\thanks{Liaoning Province Educational Committee (9808111084).}}
\author{{\small YiHuan Wei$^{1,2}$}\\
{\small $^{1}$ Institute of Theoretical Physics, Chinese Academy of Science,
 Beijing 100080, China}\\
{\small $^{2}$ Department of Physics, Bohai University, Jinzhou 121000,
 Liaoning, China}\\}
\maketitle
\begin{center}
{\bf Abstract}
\\ \vskip 0.5cm
\begin{minipage}{14cm}
\setlength{\baselineskip}{25pt plus1pt minus0pt} {\small ~~We
suggest the symmetrized Schr\"{o}dinger equation and propose a
general complex solution which is characterized by the imaginary
units $i$ and $\epsilon$. This symmetrized Schr\"{o}dinger
equation appears some interesting features. } \vskip 0.5cm
\end{minipage}
\end{center}
\vspace{0.5cm}
\setlength{\baselineskip}{25pt plus1pt minus0pt}

\section {Introduction}

Quantum physics was rapidly developed in the first three decades of last century (for a time list of some important
findings, see Ref.\cite{Rich}). The fundamental framework of quantum
mechanics was set up in 1920's \cite{Schro,Born}, (for a review, see Ref. \cite{Lal}). The interference of two
noninteracting beams of "particles" and the diffraction of single
beam of "particles" enforced us to admit its wave property. The wavefunction describing the states of the physical system
evolves according to the time-dependent Schr\"{o}dinger equation. This wave equation, which contains the first order
derivatives of the wavefunction in time, can be found by the classical one by the replacements of some classical
quantities by the corresponding operators. However, the noncommutativity of general
coordinate operator $\hat{q}$ and momentum operator $\hat{p}$ leads to ambiguities in the definition of the operator
$\hat{H}$, the wave equation (the Schr\"{o}dinger equation) is thus taken as a fundamental equation independent on a
underlying classical picture \cite{Mar}. Within quantum mechanics, the wavefunction is not a observable and what it
describes is not the wave propagating in usual three-dimensional space but a representative element of the state of the
system. The state is only a ray (or vector) in abstract $Hilbert$ space, without any classical counterpart.

In quantum mechanics, one of the postulates states that to any
self-consistent and well-defined observable in physics corresponds to an operator, the value of the observable
corresponds to the eigenvalue (expectation value) of the operator. We will use the term expectation value, instead of
eigenvalue, one can see the former will fit more to our discussions since we will deal with a symmetrized Schr\"{o}dinger
equation. The operator corresponding to the energy is the Hamiltonian $\hat{H}$. For a single particle of the
mass $\mu$, in a potential $V(\bf x)$,
$\hat{H}=\frac{\hat{\bf p}^2}{2\mu}+V(\bf x)$ with $\hat{\bf p}=-i\hbar\nabla$. The expectation value equation, i.e, the
time-dependent Schr\"{o}dinger equation yields the possible energy which the particle may have. For a free particle
the expectation value of the energy is $E=\frac{\bf p^2}{2\mu}$ with $\bf p$ the
expectation value of the momentum. This is completely consistent with the classical relationship between kinetic and
momentum.

As some preparations, let us introduce some knowledge of the complex numbers. The hyperbolic complex numbers may be
generated by the hyperbolic imaginary unit $\epsilon$ satisfying $\epsilon^2=1$. Similarly to the ordinary complex numbers,
a hyperbolic complex number takes the form \cite{Yaglom}
\begin{eqnarray}
Z_h=x+\epsilon y,
\label{eqa1}
\end{eqnarray}
where $x$ and $y$ are two real numbers, called the real and imaginary parts of the hyperbolic complex number $Z_h$. The
exponent form of the hyperbolic complex number $Z_h$ is given as follows
\begin{eqnarray}
Z_h=e^{\epsilon\theta}=\cosh\theta+\epsilon\sinh\theta,
\label{eqa2}
\end{eqnarray}
where $\theta$ is a real parameter.
For a hyperbolic complex number defined by (\ref{eqa1}), the conjugate is
\begin{eqnarray}
Z_h^\ast=x-\epsilon y.
\label{eqa3}
\end{eqnarray}
The square norm of a hyperbolic complex number is defined as
$|Z_h|^2=Z_h^\ast Z_h=x^2-y^2$. The set of hyperbolic complex
numbers has the same algebraic laws as the one of complex numbers.
It has been applied in many fields, such as, gravitational fileds
\cite{Kunsttater}, quantum group \cite{Wei}, quantum mechanics
\cite{Bracken}, {\it{etc}}. Recently, we propose so-called
$\eta$-complex number which has the two specific $\eta$-values
$\mp 1$ corresponding to the usual complex and hyperbolic complex
numbers, respectively \cite{Wei23}.

In this paper, we suggest the symmetrized Schr\"{o}dinger
equation. This equation doesn't only contain the usual
Schr\"{o}dinger equation but also shows some interesting features.
It allows such solutions that are characterized by the two
imaginary units $i$ and $\epsilon$. The three equations derived
from it have an excellent characteristic that automatically gives
the constraint on the function ${\mathcal{A}}={\mathcal{A}}(x)$
and admits a more general function
${\mathcal{S}}={\mathcal{S}}(x,t)$. This paper is organized as
follows. In Sec.II we discuss in detailed the equations derived
from the symmetrized Schr\"{o}dinger equation and gives some
special solutions. In Sec.III we give a simple comparison between
the usual Schr\"{o}dinger equation and the symmetrized
Schr\"{o}dinger equation, and analyze the relation between kinetic
energy and momentum of free particle in our framework.

\section {The Symmetrized Schr\"{o}dinger Equation}
The time-dependent Schr\"{o}dinger equation plays a fundamental role in quantum mechanics. It takes the following form
\begin{eqnarray}
i\frac \partial {\partial t}\psi=\hat{H}\psi,
\label{eqa}
\end{eqnarray}
where for simplicity we have used the unit in which $\hbar=1$, in
which the imaginary unit $i$ takes an important place. Eq.(\ref
{eqa}) has the complex solution
$\psi={\mathcal{A}}e^{i\mathcal{S}}$ or the hyperbolic complex
solution $\psi={\mathcal{A}}e^{\epsilon\mathcal{S}}$ (see
Ref.\cite{Bracken}), where ${\mathcal{A}}={\mathcal{A}}(x,t)$ and
${\mathcal{S}}={\mathcal{S}}(x,t)$ are the amplitude and phase,
respectively.

Naturally, the Schr\"{o}dinger equation may be reformulated as
\begin{eqnarray}
\frac 12(i\frac \partial {\partial t}\psi+\frac \partial {\partial t}\psi i)=\hat{H}\psi.
\label{eq1}
\end{eqnarray}
Here, let us call this equation the symmetrized Schr\"{o}dinger Equation. For the equation (\ref {eq1}) we present the
following complex solution of the form
\begin{eqnarray}
\psi={\mathcal{A}}e^{(\delta i+\sigma \epsilon)\mathcal{S}},
\label{eq2}
\end{eqnarray}
which
comprises $i$ and $\epsilon$, where $\delta$ and $\sigma$ are two constants, ${\mathcal{A}}$ and ${\mathcal{S}}$ are still
called the amplitude and phase, respectively. Noting that (see Ref.\cite{Wei})
\begin{eqnarray}
\{i,\epsilon\}=i\epsilon+\epsilon i=0,
\label{eq3}
\end{eqnarray}
by expanding $\psi$ given by Eq.(\ref {eq2}) in $\delta i+\sigma \epsilon$, then we have
\begin{eqnarray}
\psi={\mathcal{A}}[c({\mathcal{S}};\delta,\sigma)+(\delta i+\sigma\epsilon)s({\mathcal{S}};\delta,\sigma)],
\label{eq4}
\end{eqnarray}
where
\begin{eqnarray}
c({\mathcal{S}};\delta,\sigma)=\sum_{n=0}^{\infty}{\frac{(\sigma^2-\delta^2)^{n}{\mathcal{S}}^{2n}}{(2n)!}},\quad
s({\mathcal{S}};\delta,\sigma)=\sum_{n=0}^{\infty}{\frac{(\sigma^2-\delta^2)^{n}{\mathcal{S}}^{2n+1}}{(2n+1)!}},
\label{eq5}
\end{eqnarray}
which will reduce to $c({\mathcal{S}};\delta,\sigma)=cos{\mathcal{S}}$ and
$s({\mathcal{S}};\delta,\sigma)=sin{\mathcal{S}}$ for $\sigma^2-\delta^2=-1$,
$c({\mathcal{S}};\delta,\sigma)=cosh{\mathcal{S}}$ and
$s({\mathcal{S}};\delta,\sigma)=sinh{\mathcal{S}}$ for $\sigma^2-\delta^2=1$. One can find the following relation between
the function $c({\mathcal{S}};\delta,\sigma)$ and $s({\mathcal{S}};\delta,\sigma)$
\begin{eqnarray}
c^2({\mathcal{S}};\delta,\sigma)-(\sigma^2-\delta^2)s^2({\mathcal{S}};\delta,\sigma)=1,
\label{eq6}
\end{eqnarray}
which will give $cos^2{\mathcal{S}}+sin^2{\mathcal{S}}=1$ and $cosh^2{\mathcal{S}}-sinh^2{\mathcal{S}}=1$ for
$\sigma^2-\delta^2=-1,1$, respectively.
In fact, Eqs.(\ref {eq4})-(\ref {eq6}) has been given in Ref.\cite{Wei23}. From the representations of $i$ and $\epsilon$
\cite{Wei}, one can have
\begin{eqnarray}
\delta i+\sigma \epsilon=\delta\left(\begin{array}{cc}0
&-1\\
1&0\end{array}\right)+\sigma \left(\begin{array}{cc}0
&1\\
1&0\end{array}\right)=\left(\begin{array}{cc}0
&\sigma-\delta\\
\delta+\sigma&0\end{array}\right).
\label{eq7}
\end{eqnarray}
Letting $\eta=\sigma^2-\delta^2$, then Eq.(\ref {eq7}) gives a representation \cite{Wei} of $i_\eta$ with $i_\eta^2=\eta$.
The wavefunction (\ref {eq2}) can take the form $\psi={\mathcal{A}}e^{i_\eta\mathcal{S}}$ and therefore some calculations
related to the derivative of the functions $s({\mathcal{S}};\delta,\sigma)$ and $c({\mathcal{S}};\delta,\sigma)$ or
$\psi={\mathcal{A}}e^{(\delta i+\sigma\epsilon){\mathcal{S}}}$ are directly applicable, here.

According to Ref.\cite{Wei23}, we have
\begin{eqnarray}
\frac \partial {\partial q}s({\mathcal{S}};\delta,\sigma)=c({\mathcal{S}};\delta,\sigma)
\frac {\partial {\mathcal{S}}} {\partial q},\quad
\frac \partial {\partial q}c({\mathcal{S}};\delta,\sigma)= (\sigma^2-\delta^2)s({\mathcal{S}};\delta,\sigma)
\frac {\partial {\mathcal{S}}}{\partial q},
\label{eq8}
\end{eqnarray}
where $q=t, x$, or for the wavefunction $\psi={\mathcal{A}}e^{(\delta i+\sigma\epsilon){\mathcal{S}}}$,
\begin{eqnarray}
\frac \partial {\partial q}\psi=\psi\frac {\partial {ln{\mathcal{A}}}} {\partial q}+
(\delta i+\sigma\epsilon)\psi\frac {\partial {\mathcal{S}}} {\partial q}.
\label{eq9}
\end{eqnarray}
As an illustrative example of Eq.(\ref {eq2}), we will consider the case of free particle of the mass $\mu$ in
one-dimensional space. In this case,
${\mathcal{A}}=const$, ${\mathcal{S}}=px-Et$ with $p$ and $E$ two parameters, the right side ($RS$) and left side ($LS$)
of Eq.(\ref {eq1}) are
\begin{eqnarray}
RS=(\delta^2-\sigma^2)\frac{p^2}{2\mu}\psi, \quad LS=\delta E\psi.
\label{eq10}
\end{eqnarray}
Equating $RS$ to $LS$ in (\ref {eq10}) gives
\begin{eqnarray}
E=\frac{\delta^2-\sigma^2}\delta\frac{p^2}{2\mu}.
\label{eq11}
\end{eqnarray}
This equation shows the relation between the parameter $E$ and $p$. When $\frac{\delta^2-\sigma^2}\delta=1$ (two special
solutions being $(\delta=1,\sigma=0)$ and $(\delta=-1,\sigma=\pm \sqrt{2})$), it is
compatible with the classical relation between kinetic energy and momentum.

In order to derive the more general equations stemming from the
symmetrized Schr\"{o}dinger equation with the wavefunction (\ref
{eq2}), let us return to the case of three-dimensional spaces.
Writing ${\mathcal{A}}={\mathcal{A}}({\bf x},t)$ and
${\mathcal{S}}={\mathcal{S}}({\bf x},t)$, where ${\bf x}=(x_i,
i=1,2,3)$, and noting (\ref{eq3}), then the symmetrized
Schr\"{o}dinger equation (\ref {eq1}) yields
\begin{eqnarray}
\delta {\mathcal{A}} \dot{\mathcal{S}}+i\dot{\mathcal{A}}=
-\frac1{2\mu}[\nabla^2{\mathcal{A}}+(\delta i+\sigma \epsilon)({\mathcal{A}}\nabla^2{\mathcal{S}}
+2\nabla{\mathcal{A}}\cdot\nabla{\mathcal{S}})
+(\delta i+\sigma \epsilon)^2{\mathcal{A}}\nabla{\mathcal{S}}\cdot\nabla{\mathcal{S}}]+V{\mathcal{A}}.
\label{eq12}
\end{eqnarray}
where $V=V({\bf x})$ is a potential function. Since $1$, $i$ and $\epsilon$ are three independent algebraic generators,
(\ref {eq12}) will yield the three independent equations
\begin{eqnarray}
\nabla^2{\mathcal{A}}
+(\sigma^2-\delta^2){\mathcal{A}}\nabla{\mathcal{S}}\cdot\nabla{\mathcal{S}}
+2\mu\delta {\mathcal{A}} \dot{\mathcal{S}}-2\mu V{\mathcal{A}}=0,
\label{eq13}
\end{eqnarray}
\begin{eqnarray}
\dot{\mathcal{A}}=-\frac1{2\mu}\delta({\mathcal{A}}\nabla^2{\mathcal{S}}+2\nabla{\mathcal{A}}\cdot\nabla{\mathcal{S}}),
\label{eq14}
\end{eqnarray}
\begin{eqnarray}
{\mathcal{A}}\nabla^2{\mathcal{S}}+2\nabla{\mathcal{A}}\cdot\nabla{\mathcal{S}}=0,
\label{eq15}
\end{eqnarray}
where $\dot{\mathcal{A}}$ denotes the derivatives with respect to
time. The last of these three equations is the same as (12b) in
Ref.\cite{Bracken}, and the first one will reduce to (12a) when
$\sigma^2-\delta^2=1$ for $\dot{{\mathcal{S}}}=0$. There is also
an additional equation (\ref {eq14}), which prescribes the
derivative of ${\mathcal{A}}$ with respect to time. Clearly,
combining the second and third one leads to $\dot{\mathcal{A}}=0$.

In the case of one-dimensional space, the system of equations (\ref {eq13})-(\ref {eq15}) is reduced to
\begin{eqnarray}
{\mathcal{A}}^{\prime\prime}+(\sigma^2-\delta^2){\mathcal{A}}{{\mathcal{S}}^{\prime}}^2+
2\mu\delta {\mathcal{A}} \dot{\mathcal{S}}-2\mu V{\mathcal{A}}=0,
\label{eq16}
\end{eqnarray}
\begin{eqnarray}
{\mathcal{A}}{\mathcal{S}}^{\prime\prime}+2{\mathcal{A}}^{\prime}{{\mathcal{S}}^{\prime}}=0,
\label{eq17}
\end{eqnarray}
with ${\mathcal{A}}={\mathcal{A}}(x)$ which is the consequence of
Eqs.(\ref {eq14}) and (\ref {eq15}). For the cases of $\delta=1$,
$\sigma=0$ and $\sigma=1$, $\delta=0$, Eq.(\ref {eq16})
corresponds to the usual complex and hyperbolic complex situations
\cite{Bracken}, respectively. Here, we will focus on the third
special case $\sigma^2-\delta^2=0$. In this case, Eq.(\ref {eq16})
is further reduced to
\begin{eqnarray}
{\mathcal{A}}^{\prime\prime}+2\mu\delta {\mathcal{A}} \dot{\mathcal{S}}-2\mu V{\mathcal{A}}=0.
\label{eq18}
\end{eqnarray}

First, let ${\mathcal{S}}$ be only a function of space
${\mathcal{S}}={\mathcal{S}}(x)$. In this case, Eq.(\ref {eq18})
is
\begin{eqnarray}
{\mathcal{A}}^{\prime\prime}-2\mu V{\mathcal{A}}=0.
\label{eq19}
\end{eqnarray}
Although this equation take a simple form and can familiar to us we still want to show some representative examples,
such as, the constant
potential, the inverse quadratic potential, the quadratic potential, since we are considering a
special case of $\sigma^2-\delta^2=0$.

For the constant potential $V=-\eta V_0$ with $V_0>0$ and $\eta=\pm 1$, the solution of Eqs.(\ref {eq19}) and (\ref {eq17})
is given by
\begin{eqnarray}
{\mathcal{A}}=C_1c(kx;\eta)+C_2, \quad {\mathcal{S}}=B_1\int {{\mathcal{A}}^{-2}}dx+B_2,
\label{eq20}
\end{eqnarray}
with $k=(2\mu V_0)^{1/2}$, where $C_1$, $C_2$, $B_1$ and $B_2$ are integrating constants. For the $C_2=0$ case,
${\mathcal{S}}=B_1C_1^{-2}k^{-1}\frac{s(kx;\eta)}{c(kx;\eta)}+B_2$.

For the potential $V=V_0x^{-2}$ with $V_0$ a constant, the solution is
\begin{eqnarray}
{\mathcal{A}}=Cx^n, \quad {\mathcal{S}}=B_1x^{-2n+1}+B_2,
\label{eq21}
\end{eqnarray}
where $C$ is a constant and $n$ is given by $n^2-n-2\mu V_0=0$ with $V_0>-\frac{1}{8\mu}$.

For the potential $V=\frac {1}{2\mu}\sum_{n=0}^2V_nx^n$ with $V_n$
some constants, $V_2>0$ and $V_0$ being determined by
$V_0=\frac{V_1^2}{4V_2}\pm \sqrt{V_2}$, the solution is obtained
by
\begin{eqnarray}
{\mathcal{A}}=e^f, \quad f=\sum_{m=0}^2C_mx^m, \quad {\mathcal{S}}=B_1\int {{\mathcal{A}}^{-2}}dx+B_2
\label{eq22}
\end{eqnarray}
where $C_2=\pm\frac12\sqrt{V_2}$, $C_1=\pm\frac{V_1}{2\sqrt{V_2}}$
and $C_0$ is an arbitrary constant. Performing the integral of
${\mathcal{S}}$ in (\ref {eq22}) gives
${\mathcal{S}}=-\frac{\sqrt{\pi}B_1}{2\sqrt{2C_2}}e^{\frac{C_1^2}{2C_2}-2C_0}erf(\sqrt{2C_2}(x+\frac{C_1}{2C_2}))+B_2$
with $C_2>0$, where $erf(x)$ is the error function which takes on
values on the range $-1$ and $1$.

Subsequently, we consider the situation of
${\mathcal{S}}={\mathcal{S}}(t)$. In this case, Eq.(\ref {eq17})
is automatically satisfied, and we need only deal with Eq.(\ref
{eq18}). Letting $\dot{{\mathcal{S}}}=E=E(t)$, then from (\ref
{eq18}) we have
\begin{eqnarray}
{\mathcal{A}}^{\prime\prime}+2\mu(\delta E-V){\mathcal{A}}=0.
\label{eq25}
\end{eqnarray}
Noting that ${\mathcal{A}}={\mathcal{A}}(x)$, one should require that $\delta E-V$ is only a function of
space. Whether $E$ is time-dependent or not will depend on whether a potential $V$ is or not. Defining
$\widetilde{E}=\widetilde{E}(x)=-\delta E+V$, then Eq.(\ref {eq25}) takes precisely the form of Eq.(\ref {eq19}) and
therefore has the solutions that ${\mathcal{A}}$'s are given by Eqs.(\ref {eq20})-(\ref {eq22}) for
$\widetilde{E}=-\eta V_0$, $V_0x^{-2}$, $\sum_{n=0}^2V_nx^n$, which correspond to the potentials
$V=-\eta V_0-\delta E$, $V_0x^{-2}-\delta E$, $\sum_{n=0}^2V_nx^n-\delta E$, respectively.

Now, let us turn to Eqs. (\ref{eq16}) and (\ref{eq17}) with an
arbitrary $\sigma^2-\delta^2$. We have a concise method of solving
these two equations. In fact, there always is the following
relation between ${\mathcal{S}}$ and ${\mathcal{A}}$
\begin{eqnarray}
{\mathcal{S}}=B_1\int{{\mathcal{A}}^{-2}}dx+B_2
\label{eq29}
\end{eqnarray}
or
${\mathcal{S}}^{\prime}=B_1{\mathcal{A}}^{-2}$. For the $\dot{\mathcal{S}}=0$ case, putting Eq.(\ref{eq29}) into
(\ref{eq16}) yields
\begin{eqnarray}
{\mathcal{A}}^{\prime\prime}+(\sigma^2-\delta^2)B_1^2{\mathcal{A}}^{-3}-2\mu V{\mathcal{A}}=0.
\label{eq30}
\end{eqnarray}
It is easy to check that Eq.$(\ref{eq30})$ have such solutions that take the forms of these given in Table I in
\cite{Bracken}.

Here, we show a solution of Eq.(\ref{eq30}) as follows
\begin{eqnarray}
{\mathcal{A}}=Bx^{-\frac{n-1}{2}}e^{k_nx^n}, \quad {\mathcal{S}}=-\frac{B_1}{2nk_nB^2}e^{-2k_nx^n}+B_2,
\label{eq31}
\end{eqnarray}
with $B$ and $k_n$ some constants, which corresponds to the
following potential
\begin{eqnarray}
V=\frac 1{2\mu}[\frac {n^2-1}{4}x^{-2}+n^2k_n^2x^{2(n-1)}+(\sigma^2-\delta^2)B_1^2B^{-4}x^{2(n-1)}e^{-4k_nx^n}].
\label{eq32}
\end{eqnarray}
For $\sigma^2-\delta^2=1$ and $n=1$, the solution (\ref{eq31})
give the third solution in Table I, where we find the constant $C$
appearing in the solution with $V=Ce^{4kx}$ \cite{Bracken} should
be constrained to the range $C\geq0$.

\section{Discussions}
The original Schr\"{o}dinger equation has no such a solution as (\ref {eq2}) with nonzero $\delta$ and $\sigma$, thus the
symmetrized Schr\"{o}dinger equation should be a nontrivial generalization. Our solution (\ref{eq2}) has taken a more
general form, which contains both the usual complex solution and hyperbolic complex
solution \cite{Bracken} and the other special case which comprises two complex numbers, i.e., the solution given in terms
of the mixture of two complex numbers.

Here, let us proceed with our analysis of the case
$\sigma^2-\delta^2=0$. According to (\ref {eq5}), for
$\sigma^2-\delta^2=0$ we have
$s({\mathcal{S}};\delta,\sigma)=\mathcal{S}$ and
$c({\mathcal{S}};\delta,\sigma)=1$ and therefore the wavefunction
(\ref {eq4}) takes the form
$\psi={\mathcal{A}}(1+i_0{\mathcal{S}})$ or equivalently
$\psi={\mathcal{A}}e^{i_0{\mathcal{S}}}$, where $i_0\neq 0$ the
representation of which may be given from (\ref {eq7}) by letting
$\sigma=\pm \delta$. The conjugate of $\psi$ is
$\psi^\ast={\mathcal{A}}e^{-i_0{\mathcal{S}}}={\mathcal{A}}(1-i_0{\mathcal{S}})$
and the square norm $|\psi|^2=\psi^\ast \psi={\mathcal{A}}^2$, the
definitions of the conjugate and norm still holding valid in this
special case.

For $\delta=0$ and $\sigma=1$, the solution (\ref {eq2}) may be formally reduced to that of \cite{Bracken} but
it allows for the function ${\mathcal{S}}$ to take a more general form ${\mathcal{S}}={\mathcal{S}}(x,t)$. For
the constant ${\mathcal{A}}$ and ${\mathcal{S}}={\mathcal{S}}(x,t)$, Eq.(\ref {eq17}) will yield
${\mathcal{S}}={\mathcal{S}}_1 x+{\mathcal{S}}_2$, where ${\mathcal{S}}_1={\mathcal{S}}_1(t)$,
${\mathcal{S}}_2={\mathcal{S}}_2(t)$. Putting it into Eq.(\ref {eq16}) gives
\begin{eqnarray}
(\sigma^2-\delta^2){\mathcal{S}}_1^2+2\mu \delta(\dot{{\mathcal{S}}_1}x+\dot{{\mathcal{S}}_2})-2\mu V=0.
\label{eq33}
\end{eqnarray}
Letting ${\mathcal{S}}_1=p=p_0+kt$ with $p_0$ and $k$ two
constants and ${\mathcal{S}}_2=Et$ with $E$ also a constant, then
(\ref {eq33}) reduces to
$E+\dot{E}t+kx-\delta^{-1}V=\frac{(\delta^2-\sigma^2)p^2}{2\mu
\delta}$, where $V$ is required to be a linear potential $V=\delta
kx$, such as, the gravitational potential energy of a particle and
the potential energy of an electric charge in an electric field of
a uniform strength, which gives (\ref {eq11}) for $k=0$. As
mentioned above, it is only classically acceptable for
$\frac{\delta^2-\sigma^2}{\delta}=1$ with $p$ and $E$ being
interpreted as momentum and kinetic energy, respectively.

In fact, as well known, not each effect of quantum mechanics can
have the classical counterpart. So, the other cases of (\ref
{eq11}) could probably be physically meaningful results. In the
following, we will give some discussions of the relation between
the kinetic energy and momentum of a particle. The purpose of
doing this is to attempt to reveal some possible quantum effects
that are banned by classical mechanics.  Let now us give a simple
analysis of Eq. (\ref {eq11}).  The two special cases of it, $E=0$
and $E<0$, are interesting, which correspond to
$\delta^2-\sigma^2=0$ and $\frac{\delta^2-\sigma^2}{\delta}<0$
with nonzero $\delta$, respectively, where $p$ is assumed to be
real, nonzero. They exceed the range of classical mechanics and
also cannot be obtained from the usual Schr\"{o}dinger equation
but are the consequences of the symmetrized Schr\"{o}dinger
equation indeed. The latter of these two cases may imply a phantom
particle of mass $\mu$, momentum $p$. The phantom energy is said
to be a negative kinetic energy, it is allowed in condensate
physics, cosmology and the current astronomical observation
\cite{Mari,Robert,RobertC}. Since the phantom energy may exist in
nature, a particle of the mass $\mu$, nonzero momentum $p$ and
zero kinetic energy should also be acceptable. Our discussions of
the relation between kinetic energy and momentum can be to some
extent consistent with the conjecture \cite{Jirari,Kroger} that
for a given classical action $S=\int dt\frac{m}{2}\dot{x}^2-V(x)$
with $m$ mass and $x$ the position coordinate of a particle there
is a quantum action $S=\int
dt\frac{\widetilde{m}}{2}\dot{x}^2-\widetilde{V}(x)$ in the path
integral approach, where the kinetic energy and potential in the
quantum action take the modified forms. In fact, this has also
suggested that the relation between kinetic energy and momentum in
quantum mechanics needn't to be equivalent completely to that in
classical mechanics.

\end{document}